\documentclass[twocolumn,showkeys]{revtex4-2}

\newcommand{\lvec}{\mathbf{l}}
\newcommand{\svec}{\mathbf{s}}

\usepackage{todonotes}

\begin{document}

\title{High-throughput design of cultured tissue moulds using a biophysical model}

\author{James P. Hague}
\affiliation{School of Physical Sciences, The Open University, Milton Keynes, MK7 6AA}

\author{Allison E. Andrews}
\affiliation{School of Physical Sciences, The Open University, Milton Keynes, MK7 6AA}

\author{Hugh Dickinson}
\affiliation{School of Physical Sciences, The Open University, Milton Keynes, MK7 6AA}

\begin{abstract}

The technique presented here identifies tethered mould designs, optimised for growing cultured tissue with very highly-aligned cells. It is based on a microscopic biophysical model for polarised cellular hydrogels. There is an unmet need for tools to assist mould and scaffold designs for the growth of cultured tissues with bespoke cell organisations, that can be used in applications such as regenerative medicine, drug screening and cultured meat. High-throughput biophysical calculations were made for a wide variety of computer-generated moulds, with cell-matrix interactions and tissue-scale forces simulated using a contractile-network dipole-orientation model. Elongated moulds with central broadening and one of the following tethering strategies are found to lead to highly-aligned cells: (1)  tethers placed within the bilateral protrusions resulting from an indentation on the short edge, to guide alignment (2) tethers placed within a single vertex to shrink the available space for misalignment. As such, proof-of-concept has been shown for mould and tethered scaffold design based on a recently developed biophysical model. The approach is applicable to a broad range of cell types that align in tissues and is extensible for 3D scaffolds.

\end{abstract}

\date{\today}


\maketitle

\section{Introduction}

It is both difficult and time consuming to design moulds and scaffolds to make artificial tissue with realistic characteristics, and new tools to speed up this design process are needed. Cultured tissues mimic the 3D structures in real tissues and have applications as regenerative medicine \cite{bajaj2014a}, pharmacological assays  \cite{weinhart2019a,jensen2018a}, cultured meats \cite{benarye2019a}, and for the study of fundamental biology. Several growth strategies for 3D engineered tissue are available \cite{bajaj2014a}. Tethered, cell-laden hydrogels can be a convenient way to make such tissues and have potential for applications in high-throughput screening (e.g. muscle \cite{capel2019a,wragg2019a}), drug testing (e.g. tendons \cite{garvin2003a}) and growing small volumes of tissue for regenerative medicine (e.g. neural tissue \cite{georgiou2013} and cornea \cite{mukhey2018}). 

Depending on the application, there may be several different objectives for a mould or scaffold design. It may be desired to maximise the overall alignment of the tissue and minimise the tension in the sample (as tension can cause challenges such as breakage if it is necessary to remove the hydrogel from the mould). When growing highly aligned or polarised tissues using tethering, there is often a misaligned region shaped like a Greek $\Delta$ close to the tethers (see e.g. Ref. \cite{eastwood1998}). Depending on the application, minimising the size of this $\Delta$-region can be important. There may also be constraints on size and shape of the mould related to materials or growth media. This can make the design process challenging. In this paper, we carry out a proof-of-concept study to show how the use of biophysical simulations can help to guide design considerations.

In this paper, we use biophysical simulations of cell-matrix interactions as a way to test large numbers of mould designs in parallel and receive immediate information regarding cell organisation, tension, tissue density and other properties \cite{hague2019a,andrews2023}. The extracellular matrix (ECM) provides structure to tissue and is important for tissue development, maintenance and repair \cite{kular2014a}. The biophysical contractile-network-dipole-orientation (CONDOR) model used here is a microscopic approach that simulates cell-matrix interactions in tissues (a key feature of cell behaviour in tissues \cite{kular2014a} that leads to self-organisation). Using this method, the self-organisation of cells in tissue sizes of up to a few mm can be simulated relatively quickly. When combined with high-performance computing, hundreds of mould designs can be tested and analysed automatically within a few days. This allows for the pre-screening of designs before the lengthy process of growth and microscopy in the lab. CONDOR has been shown to accurately predict how cell-matrix interactions drive shaping and self-organisation of tethered cellular hydrogels of glial cells \cite{hague2019a}, and similar self-organisation is found in hydrogels of skeletal muscle \cite{wragg2019a}, tendons \cite{garvin2003a}, fibroblasts \cite{eastwood1998} and corneal tissue \cite{mukhey2018}.

The goal of this paper is to show proof-of-concept for the use of the CONDOR approach to identify the best mould shapes and arrangements of tethers for cultured tissue growth with specific characteristics (such as high levels of cellular alignment and low tension). In Section \ref{sec:methodology}, we introduce methods for the automated design of moulds and tethers for tissue growth and briefly review the CONDOR model. In Section \ref{sec:results} we identify moulds for growing cultured tissues with very high levels of cellular alignment. A discussion is found in Section \ref{sec:discussion}. Conclusions and summary can be found in Section \ref{sec:conclusions}.

\section{Methodology}
\label{sec:methodology}

\subsection{Mould generation}

We have developed a procedure to automatically generate plausible tethered mould designs for highly aligned tissues \cite{andrews2023}. In this paper, we use a subset of these moulds. We begin by randomly generating a set of 3 to 6 points on a 2D canvas with $x$ and $y$ coordinates in the range $\left[0,1\right]$. These points are then used to define a simple polygon or convex hull. We reject any shapes with a total area of less than half the canvas area. We generate 3 circular tethers at random positions within this polygon, each with a radius between $0.02$ and $0.065$. A single random position within this polygon is set as the origin for axes of symmetry. A shape mirrored along both axes is produced by firstly duplicating the initial polygon and mirroring along one axis, then duplicating both the initial and new polygon and mirroring on the perpendicular axis. These are combined into a single polygon that is symmetric on both $x$ and $y$ axes. Tethers are duplicated and mirrored in the same fashion.

\begin{figure*}
\includegraphics[width=0.9\textwidth]{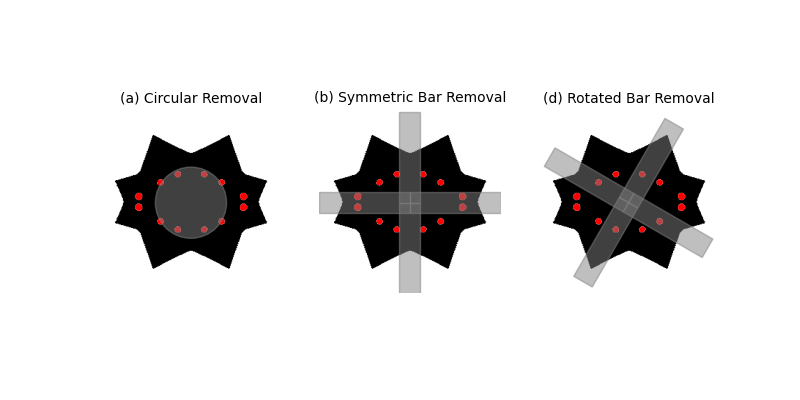}
\caption{Examples of the different types of exclusion areas used for tether removal, including circular removal (a), symmetric bar removal (b), and rotated bar removal (d).}
\label{fig:tether_removal_examples}
\end{figure*}

After duplication, tethers are pruned. This modifies the distribution and reduces the total number of tethers present within the final mould. Several types of exclusion area for tethers are defined: (a) For 45\% of cases, a circular exclusion area is placed in the centre of the mould with a diameter that is selected randomly to be between 0.2 and 0.8 times the maximum mould width. (b) For 45\% of cases, exclusion areas are defined as bars centred on one or both of the axes of symmetry with width between 0.1 and 0.4 of the maximum mould width. (c) For 5\% of  moulds we use a variation of the bar removal regime in which the perpendicular bars are rotated about the symmetry origin by a random angle; the final regime has the potential of producing moulds with rotationally symmetric tether arrangements rather than the mirror symmetry of the mould.  (d) For 5\% of moulds we omit tether removal. Examples of the exclusion area defined in removal regimes (a), (b) and (c) are shown in Figure~\ref{fig:tether_removal_examples}. If, after this process, any mould has less than 4 tethers remaining, it is discarded. 

The Python Shapely package (version 2.0.1) is used to refine the design \cite{shapely2023}. The Shapely function `simplify' is applied to the object with tolerance 0.05 to remove very intricate vertices with higher resolution than the simulations, with the caveat of minor deviations from symmetry occurring for some generated moulds; this is due to the function not always selecting corresponding mirrored vertices for removal and the biophysical simulations reflect this deviation. 

We use the `buffer' method from Shapely to create curved edges of both concave and convex type in the mould design. Applying a negative buffer value followed by a positive value (i.e. an antibuffer followed by a buffer) creates a convex curved structure where outer corners exist. Likewise the opposite process with a positive buffer value followed by a negative value generates a concave structure with inner corners. Both of these processes can be applied independently without affecting the structures already generated by the other. Our mould generation has a 25\% probability of applying the convex rounding, and separately a 75\% probability of applying concave rounding regardless of whether the convex rounding is applied; this subsequently results in approximately $18.75$\% of cases having both types of rounding applied, as well as another $18.75$\% having neither. Examples of moulds generated in this way can be seen throughout the paper. 

\subsection{CONDOR simulation}

We simulate the self-organisation of tissues grown within these mould shapes using the contractile network dipole orientation (CONDOR) model \cite{hague2019a}, which we briefly review here. The model comprises a contractile network of bonds following Hooke's law, with cells sitting at the end of bonds. Cell-matrix interactions are incorporated by modifying each bond's equilibrium length according to the relative bond-to-cell orientation. The form of the model is,
\begin{equation} 
E = \sum_{i<j} \frac{\kappa_{0}\bar{\kappa}_{ij}}{2}\left( |\lvec_{ij}| - l'_{ij}\right)^2
\end{equation}
where the adjustment to the bond length due to cell-matrix interactions is,
\begin{equation}
l'_{ij} = l_{0}\left(1-\frac{\Delta_{\rm CM}}{2}\left(2-|\hat{\lvec}_{ij}\cdot\svec_{i}|^2 -|\hat{\lvec}_{ij}\cdot\svec_{j}|^2\right) \right).
\end{equation}
Here, $E$ represents the total energy of the system of cells and force dipoles. $\Delta_{\rm CM}(=0.3)$ represents the cell-matrix interaction strength (note that the model parameter $\Delta_{\rm CM}$ should not be confused with the $\Delta$ region). $l_{0}$ is the equilibrium distance of the bonds between cells in the absence of cell-matrix interactions.  We nominally set $l_{0}=40\mu\mathrm{m}$, although we note the model is scale independent and results are given in terms of dimensionless quantities.  $\lvec_{ij}$ is the displacement between cells $i$ and $j$. The unitary vector $\svec_{i}\equiv(s_x,s_y,s_z)$ represents the orientation of cell $i$. $\kappa_{0}(=1)$ is the nearest-neighbour spring constant, and only sets the energy scale (i.e. results are independent of $\kappa_{0}$). We run simulations with cells arranged on a face-centre-cubic (FCC) lattice to ensure shear modulus in the network. To prevent the whole lattice from folding about a plane, we include more distant bonds in the contractile network so that folds cost a high energy due to compression in the longer range bonds. The dimensionless spring constant, $\bar{\kappa}_{ij}$, has the specific values $\bar{\kappa}_{\rm NN}=1$,  $\bar{\kappa}_{\rm NNN} = 0.4$, $\bar{\kappa}_{\rm NNNN} = 0.2$ and $\bar{\kappa}_{\rm NNNNN} = 0.2$ where NNN, NNNN and NNNNN represent next-nearest through next-next-next nearest neighbours respectively.

The mould is scaled so its area is exactly 40\% of the total simulation area of $62.5 l_0 \times 62.5 l_0$ and is given a depth, $4 l_{0}$. The lowest energy state of this model is found using simulated annealing \cite{hague2019a}. Change in position or orientation of a single cell up to a maximum distance or difference in angle is attempted with equal probability on each iteration. The initial temperature of the anneal is fixed at ${\mathcal{T}}_{\rm init}/\kappa_{0}l_{0}^2=0.0625$, a fixed number of iterations is used ($N_{\rm cells} \times 10^{6}$, where $N_{\rm cells}$ is the total number of cells in the simulation) and the ratio between initial and final anneal temperatures ${\mathcal{T}}_{\rm init}/{\mathcal{T}}_{\rm final} = 10^{7}$ (note that the anneal temperature is defined for simulated annealing only and is not the temperature of the tissue). For the mould sizes selected, $N_{\rm cells}\approx 10000$. A penalty of $10^{9}$ is applied to the energy for each cell that is situated outside the mould or inside a tether. The CONDOR simulation of each individual tethered mould in this set takes approximately 24-36 core hours. We run the simulations in parallel on a large multi-core machine allowing for rapid throughput. Further details can be found in Ref. \cite{hague2019a}.

After the anneal is complete, we determine physical properties of the simulated tissue within the moulds. Cell orientation is read directly from the model. Average cell orientation in the $x$-direction is calculated as,
\begin{equation}
    s^{2}_{x,{\rm av}} = \frac{1}{N_{\rm cells}}\sum_{i} s^{2}_{x}
\end{equation}
with similar measures in the $y$ and $z$ directions. Two measures of the tension are determined. The total of the magnitude of the tension each bond connected to an individual cell $i$: 
\begin{equation}
\tau_{i} = \sum_{j}|T_{ij}|
\label{eqn:totaltensionpercell}
\end{equation}
and the average tension per bond 
\begin{equation}
    |T|_{\rm av} = \frac{1}{N_{\rm bonds}}\sum_{i<j} |T_{ij}|
    \label{eqn:averagetensionperbond}
    \end{equation}
    where 
    \begin{equation}
        T_{ij} = \kappa_{0}\bar{\kappa}_{ij}\left(|\lvec_{ij}|-l'_{ij}\right).
        \end{equation}
and $N_{\rm bonds}$ is the total number of bonds.

\section{Results}
\label{sec:results}

\begin{figure}
\includegraphics[width=0.48\textwidth]{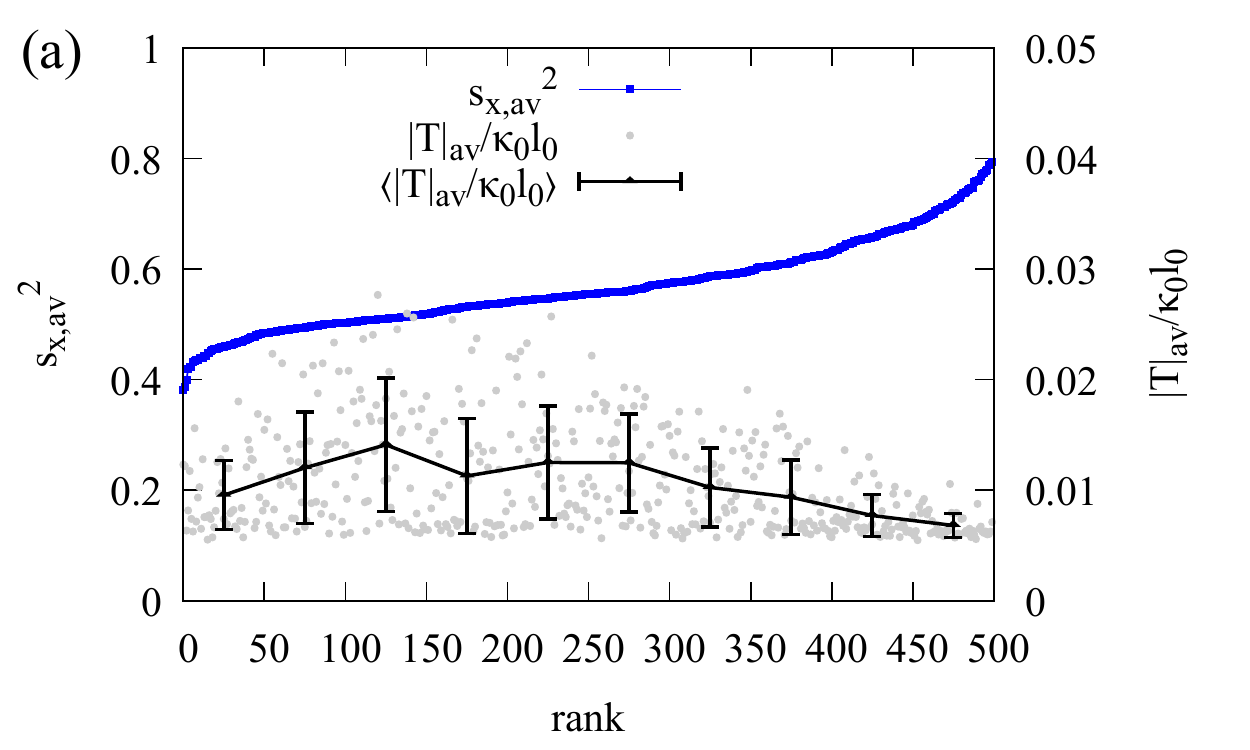}
\includegraphics[width=0.48\textwidth]{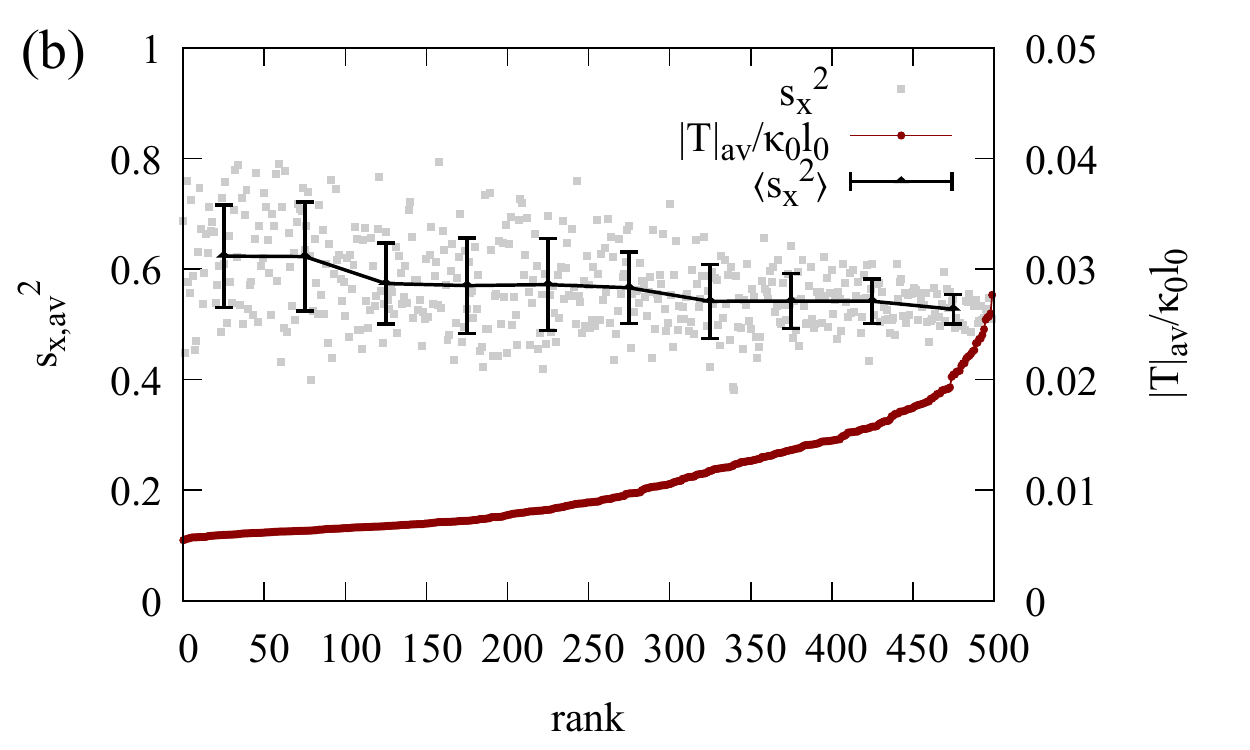}
\caption{(a) Ranked alignments show that moulds containing tissue with the highest alignments have low tension and a maximum average alignment $s_{x,\rm av}^2\sim 0.8$. Moulds with high tension tend to have poor alignment. Ranked alignments are shown as blue points. The scatter of tension for each mould is also shown in light grey, with the mean and standard deviation calculated from blocks of 50 moulds to highlight the trend in tension with alignment rank (black line with error bars). (b) Tensions sorted by rank order (red points). The scatter of alignment for each case is also shown in light gray. The moulds with the highest alignments tend to be associated with low tensions, but it is also possible for low tension moulds to have poor alignment. Moulds with the highest tensions have poor alignment.}
\label{fig:alignmenttensionrank}
\end{figure}

After simulation of the self-organisation of cells within moulds, results were ranked according to their total alignment, showing that alignments can exceed 80\% of the maximum possible alignment, and that the best aligned moulds have low tension. Specifically, we rank according to the largest $s_{x,{\rm av}}^2$. In the event that $s_{y,{\rm av}}^2$ is larger than $s_{x,{\rm av}}^2$ (which typically happens if the mould is elongated on the $y$ instead of the $x$ direction) the mould (and its contents) is rotated by 90 degrees to facilitate comparison. Fig. \ref{fig:alignmenttensionrank}(a) shows average alignments by rank order. Tension for each mould is shown as a scatter and the average tension and its standard deviation for bins of 50 closely ranked moulds is displayed with error bars to show the trend. The largest alignments in the moulds considered exceed $s_{x,{\rm av}}^2\sim 0.8$ and the lowest are around $s_{x,{\rm av}}^2\sim 0.4$ (note that for random alignments in 3D, $s_{x,{\rm av}}=1/3$). Moulds containing simulated tissue with the highest alignments have low tension. Moulds with high tension tend to have poor alignment. Overall, tension and alignment do not show a strong correlation. The trend is that the maximum possible average tension decreases with alignment. We note that it is possible to find poorly aligned moulds with low tension. There are no highly aligned moulds with high tension.

Tension is not a strong predictor of alignment, as shown by ranking by average tension (Fig. \ref{fig:alignmenttensionrank}(b)). Alignment for each mould is also shown as a scatter for comparison. The average and standard deviations of tension are shown as errorbars for bins of 50 closely ranked moulds. Again, there is no overall correlation between tension and alignment. The moulds with the highest alignments are associated with low tensions, but it is also possible for low tension moulds to have poor alignment. Moulds with the highest tensions tend to have poor alignment. The maximum possible alignment tends to descrease as tension increases. We note that the correlation between maximum possible alignment and tension is weaker than the correlation between maximum possible tension and alignment. Since the goal here is to find highly aligned moulds with low tension then ranking according to alignment is sufficient to ensure that the simulated tissue is under low tension.

\begin{figure*}
\includegraphics[width=0.9\textwidth]{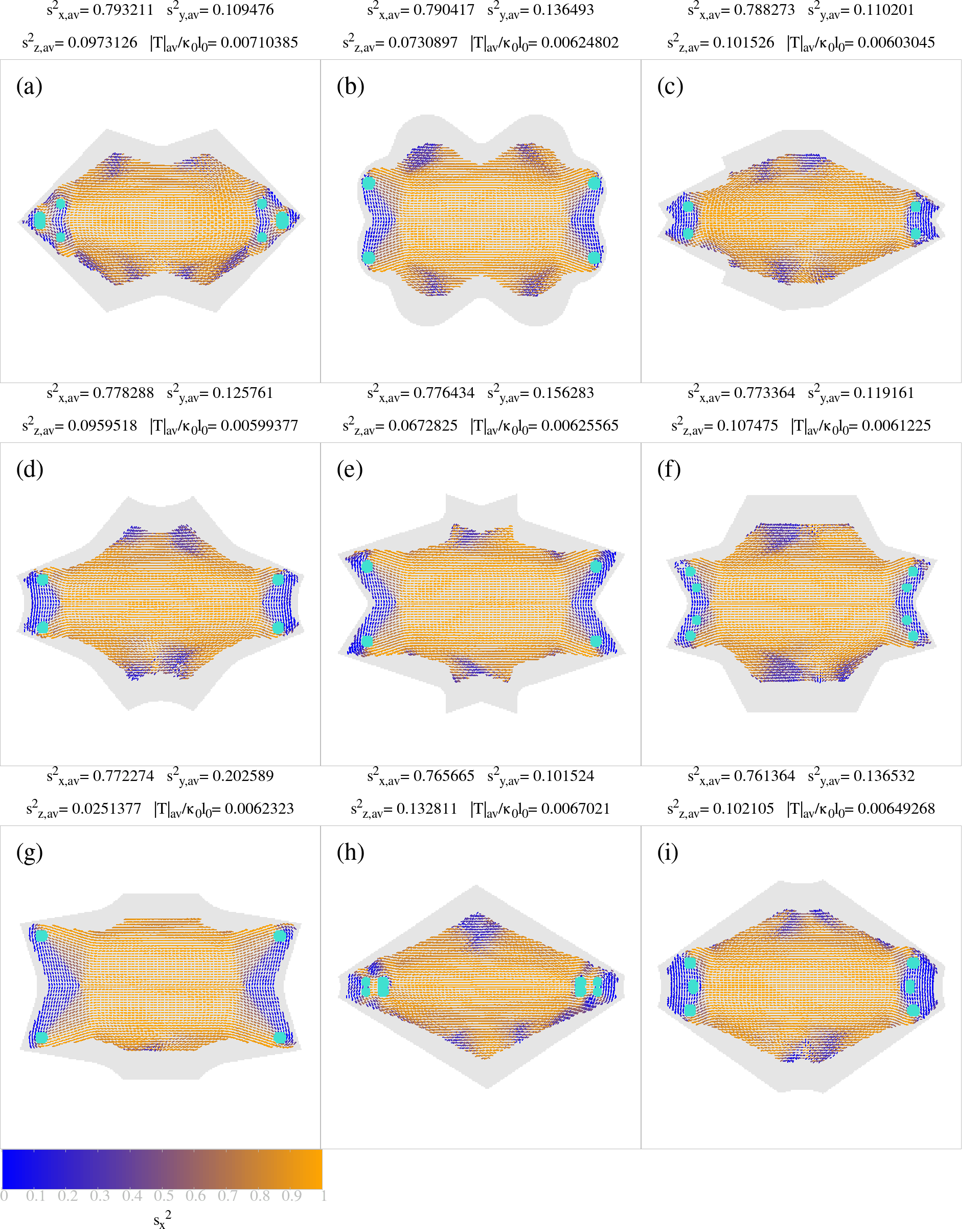}
\caption{Top 9 moulds (rank 492-500) with regard to orientation indicate two strategies for maximising alignment. Cells in the simulated tissue are highly aligned. The two strategies are: (1) indentations in the short axis act to reduce the size of the $\Delta$ region, (2) double columns of tethers placed within a single vertex reduce the $\Delta$ region. Tensions tend to be slightly higher for strategy (2) as seen in Fig. \ref{fig:bestorienttension}. In the figures, blue indicates alignment along the $y$ axis and orange along the $x$ axis. Tethers are shown in cyan and the mould area in light grey. For convenience average alignment and tension per bond for each mould are shown above the plots.}
\label{fig:bestorient}
\end{figure*}

Figure \ref{fig:bestorient} shows the best 9 moulds with regard to orientation, highlighting two strategies for maximising alignment of cells. Moulds leading to highly aligned (simulated) tissue are elongated and contain tethers located close to the short edge of the mould. The alignment is optimised by minimising the $\Delta$ region via one of two approaches: The first strategy is an indentation or concave region on the short edge of the mould, with tethers then placed within the protrusions on either side of the indentation Fig. \ref{fig:bestorient}(b,d,e,f,g). The second strategy is that multiple columns of tethers are found within a single convex vertex at the end of the short edge, Fig. \ref{fig:bestorient}(a,h). In two of the cases shown, a hybrid strategy leads to a very short edge with a small indentation Fig. \ref{fig:bestorient}(c,i). In all cases the moulds are narrowed towards the short edge and are typically much broader in the central region of the mould. This is likely to increase the proportion of aligned tissue since the smaller short edge decreases the potential area of the misaligned region. The long edge is more complicated with fewer commonalities. Several examples have an indentation towards the centre of the long edge, but protrusions to almost diamond shaped moulds also occur. Moulds with a broader centre may lead to squarer regions of polarised tissue centrally e.g. Fig. \ref{fig:bestorient}(f), but highly aligned rounded regions are also possible Fig. \ref{fig:bestorient}(i). Square or circular regions of aligned tissue may have different applications (for e.g. repair or representation of a region of cornea). Differences in total alignment between the best examples shown here are approximately 5\%.

\begin{figure*}
\includegraphics[width=0.9\textwidth]{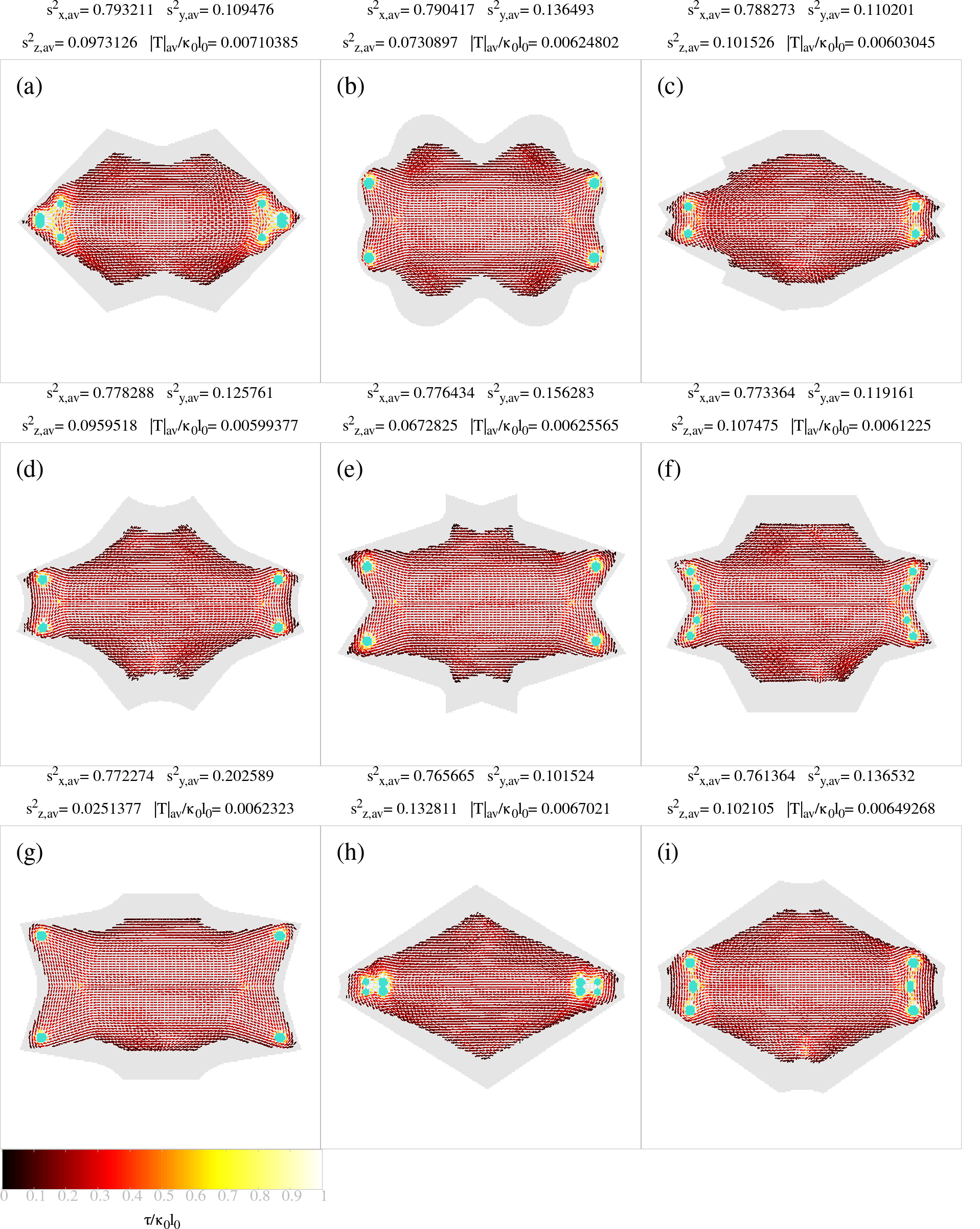}
\caption{Tensions within the top 9 moulds (ranks 492-500) ranked with regard to orientation are generally low. For strategies where two columns of tethers have reduced the $\Delta$ region, tension is high between the two columns. In the figures, black and red indicate regions of low tension, yellow and white indicate regions of high tension. Tethers are shown in cyan and the outline of the mould in grey.}
\label{fig:bestorienttension}
\end{figure*}

Fig. \ref{fig:bestorienttension} shows that the absolute value of the tension on a cell, $\tau_{i}$, is relatively small in all the highly aligned cases identified in Fig. \ref{fig:bestorient}. We define the absolute value of the tension to be the sum of the magnitudes of the tensions in all bonds associated with a single cell. Relative to tissue simulated in other moulds, the tension is smallest for highly aligned simulated tissues. Tension is highest close to tethers where there are regions of rapidly changing alignment. For moulds with multiple columns of tethers, shown in Fig. \ref{fig:bestorienttension}(a,h), tension can be high between the tethers, which may cause challenges if it is necessary to remove the tissue from the mould. Note that the average tension per bond as defined in Eq. \ref{eqn:averagetensionperbond} is shown above the plots, whereas the total tension on a cell as defined in Eq. \ref{eqn:totaltensionpercell} is shown within the plots.

\begin{figure*}
\includegraphics[width=0.9\textwidth]{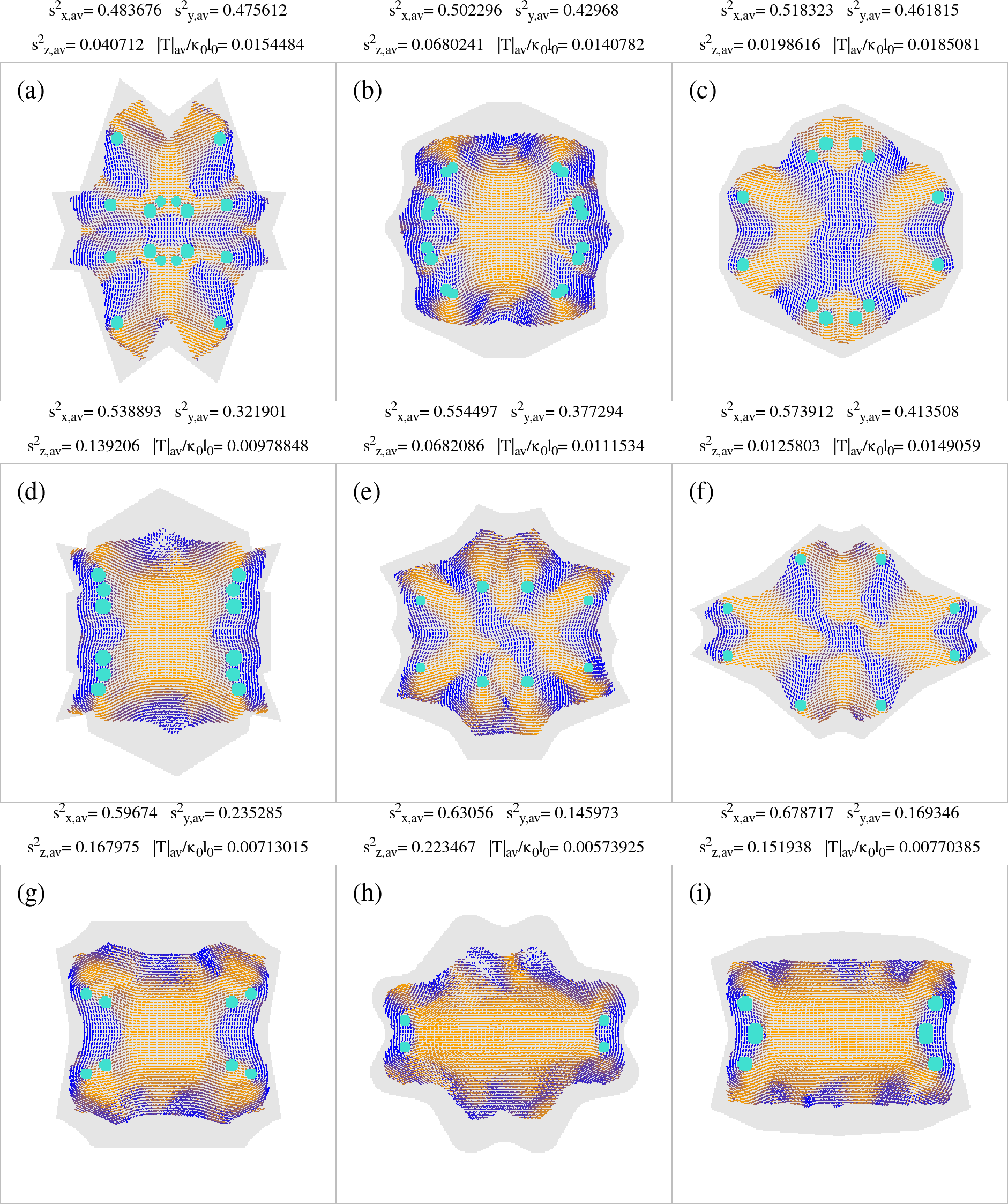}
\caption{Figure showing how alignment improves by ranking in a range of designs. Ranks 51 to 451 are shown in steps of 50. For lower ranked cases, tethers are spread across the mould and moulds do not have a long axis. Higher ranked moulds have a long axis with tethers situated closest to the short axis. Small indentations along the short axis help to reduce the size of the $\Delta$ region.}
\label{fig:rankedmoulds}
\end{figure*}

To understand why some strategies work well, it can be helpful to examine cases where alignment is predicted to be far from optimal, which occurs in e.g. moulds with excessive numbers of (or conflicting) tethers. Examples of simulated tissue ranked according to alignment in steps of 50 are shown in Fig. \ref{fig:rankedmoulds}, where the lowest ranked moulds have the worst alignment. The presence of excessive numbers of tethering points leads to regions of poorly aligned cells between the tethers. The worst aligned example in Fig. \ref{fig:rankedmoulds}(a), which is rank number 51, has tethers distributed broadly across the mould area. As alignment increases, tethers move towards the short edge of the mould. For the highest aligned example shown in this figure (Fig. \ref{fig:rankedmoulds}(i), which is rank 451) the tethers are situated directly beside the short edge of the mould and the mould is elongated along the $x$ axis.

\begin{figure*}
\includegraphics[width=0.9\textwidth]{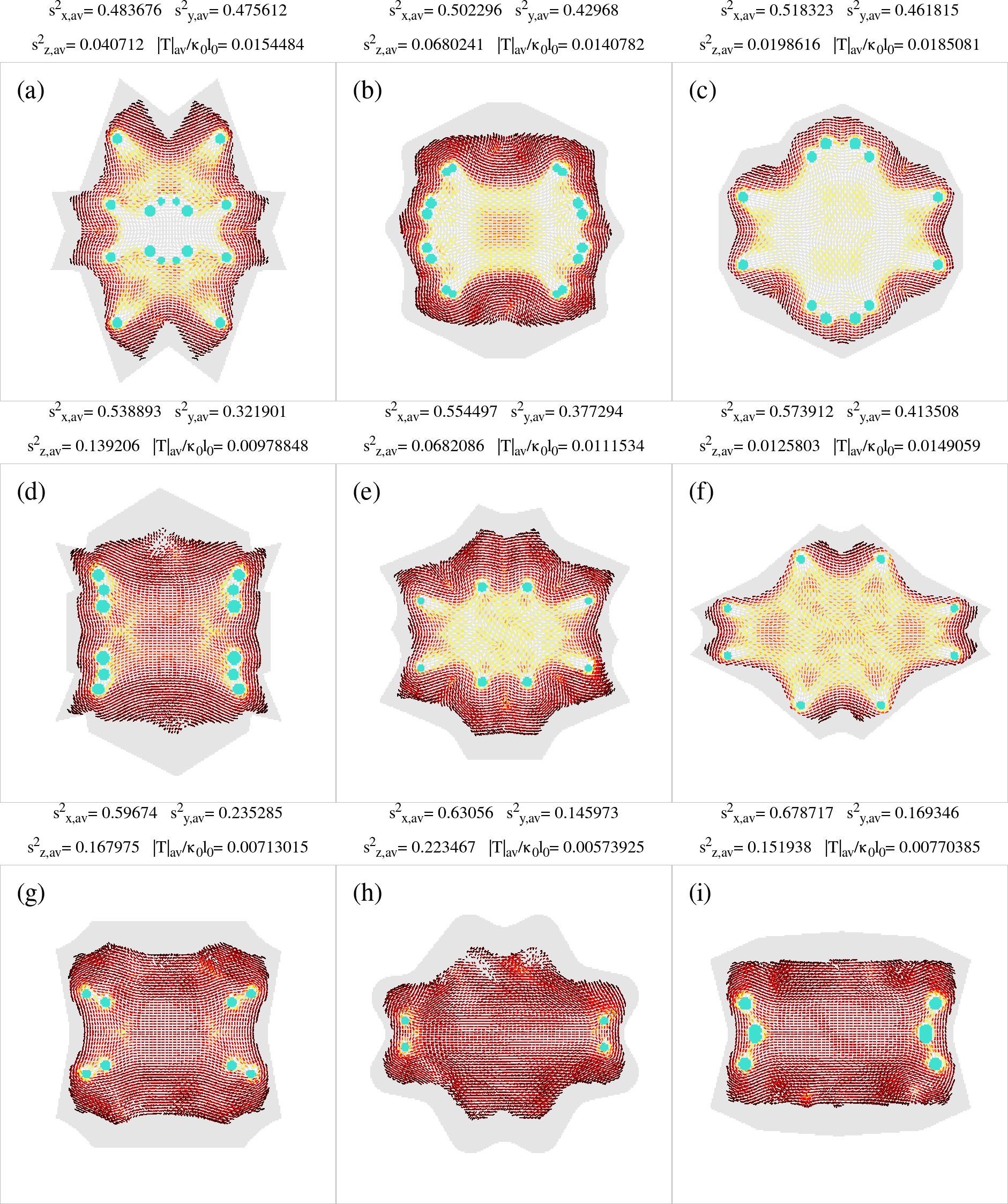}
\caption{Analysis corresponding to the designs shown in Fig. \ref{fig:rankedmoulds} demonstrates how tension is affected by tether number and positioning. Areas of high tension are associated with misaligned regions and tend to decrease with rank. Large numbers of tethers away from the short axis typically lead to high tension.}
\label{fig:rankedmouldstension}
\end{figure*}

High tension is typically associated with regions of frustration, where orientations change rapidly, which often occurs around tethers, and also in the bulk of the tissue when there are large numbers of conflicting tethers. Analysis of tension for the ranked moulds shown in Fig. \ref{fig:rankedmouldstension} highlights the causes of misaligned regions (all cases correspond to the moulds in Fig. \ref{fig:rankedmoulds}). In many moulds which are predicted to have poorly aligned cells, the tissue is typically under high tension. High tension may also occur away from tethering points if other tethers promote alignment in conflicting directions causing frustration towards the centre of the tissue. Tension is low for the best aligned cases shown in Fig. \ref{fig:rankedmouldstension}(g,h,i). In the remaining figures, we consider the effect of specific changes to mould shapes and the positioning of tethers on the alignment.

\begin{figure*}
\includegraphics[width=0.9\textwidth]{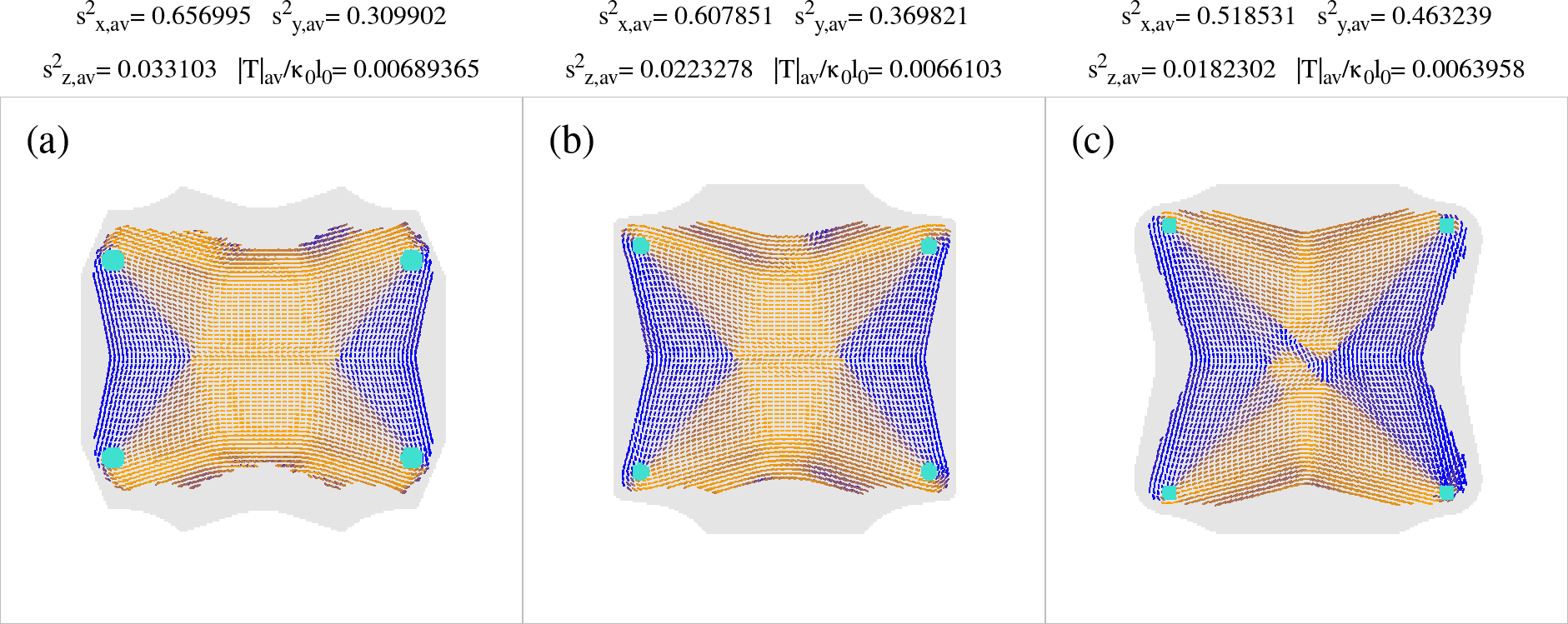}
\caption{Change in $\Delta$ region due to decrease in ratio of short to long axis. Overall the shape of the $\Delta$ region is not strongly affected by the length of the mould, but misaligned regions can dominate if the mould is too short.}
\label{fig:shortlongratio}
\end{figure*}

Elongation is important to reduce the relative size of the $\Delta$-region compared to the overall mould, but otherwise does not decrease the area of the $\Delta$ region. In Fig. \ref{fig:shortlongratio} we show examples where the ratio of short to long axes varies. For the elongated mould in Fig. \ref{fig:shortlongratio}(a), the $\Delta$ regions are of similar size but the region of aligned tissue is larger due to the anisotropy in the mould. For the almost square mould in Fig. \ref{fig:shortlongratio}(c), the $\Delta$ regions from the two axes overlap.

\begin{figure}
\includegraphics[width=0.3\textwidth]{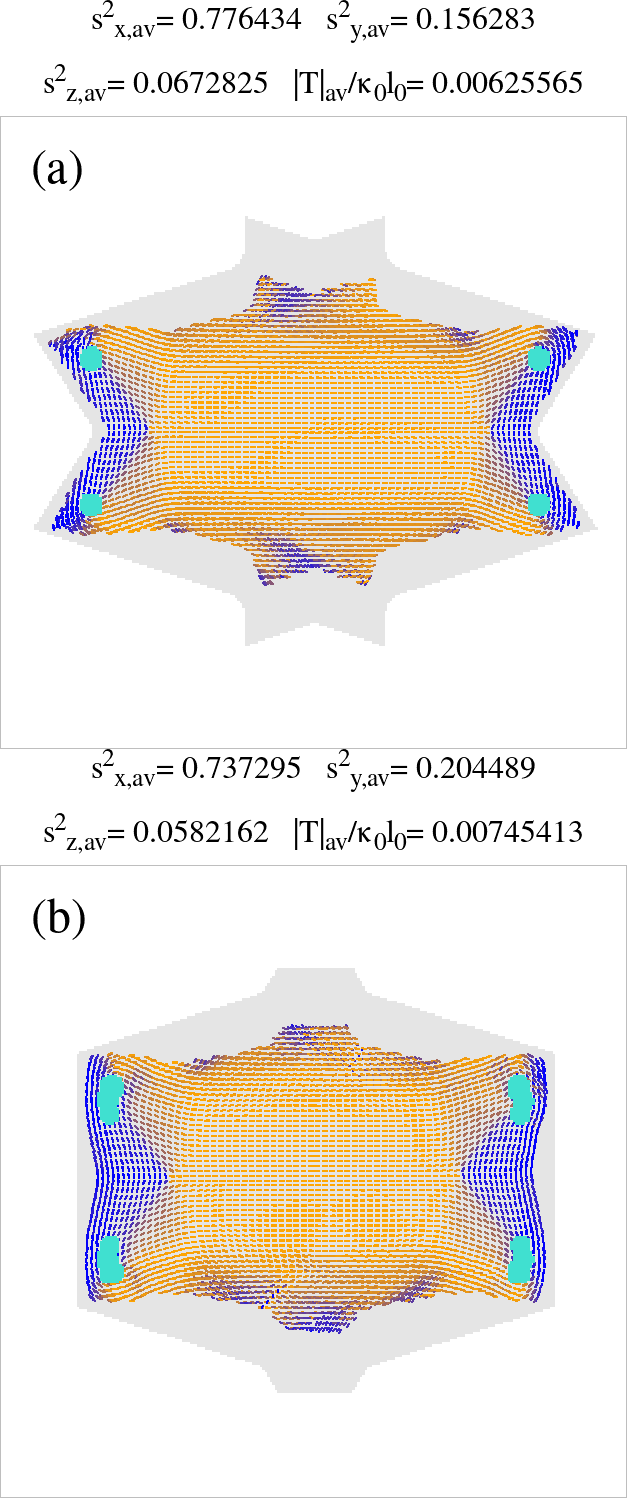}
\caption{Significant decreases in the $\Delta$ region can be made by inserting a concave region on the short axis.}
\label{fig:deltaindent}
\end{figure}

Indenting the short edge leads to a reduced $\Delta$-region size. In Fig. \ref{fig:deltaindent} we show two similar examples, one with a significant indent on the short axis and the other with a straight short edge. The indent guides the cells in the right direction, and supplants the misaligned $\Delta$ region. The tension (not shown) is low in both cases.

\begin{figure}
\includegraphics[width=0.3\textwidth]{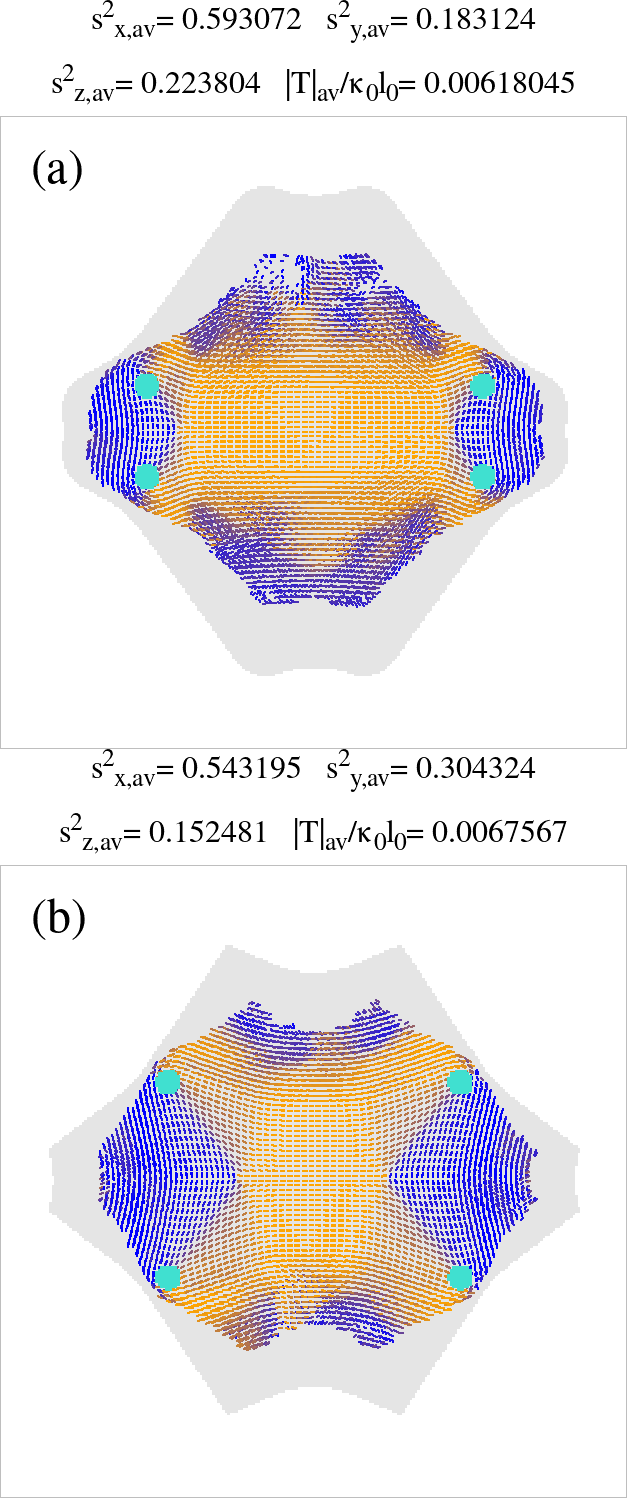}
\caption{Changes in the size of the $\Delta$ region due to positioning of tethers. Poorly positioned tethers that are not tight to the short axis vertices lead to larger $\Delta$ regions.}
\label{fig:poorplacement}
\end{figure}

Poorly positioned tethers that are not close to the short axis vertices lead to larger $\Delta$ regions, as can be seen in Fig. \ref{fig:poorplacement}. In Fig. \ref{fig:poorplacement}(a) the tethers are closer to the short edge and the $\Delta$ region is relatively smaller than in Fig. \ref{fig:poorplacement}(b), where the tethers are far from the short edge, but otherwise close to the boundary of the mould. This highlights how the positioning of tethers, as well as the shape of the mould, can be very important for creating polarised tissues.

\begin{figure*}
\includegraphics[width=0.9\textwidth]{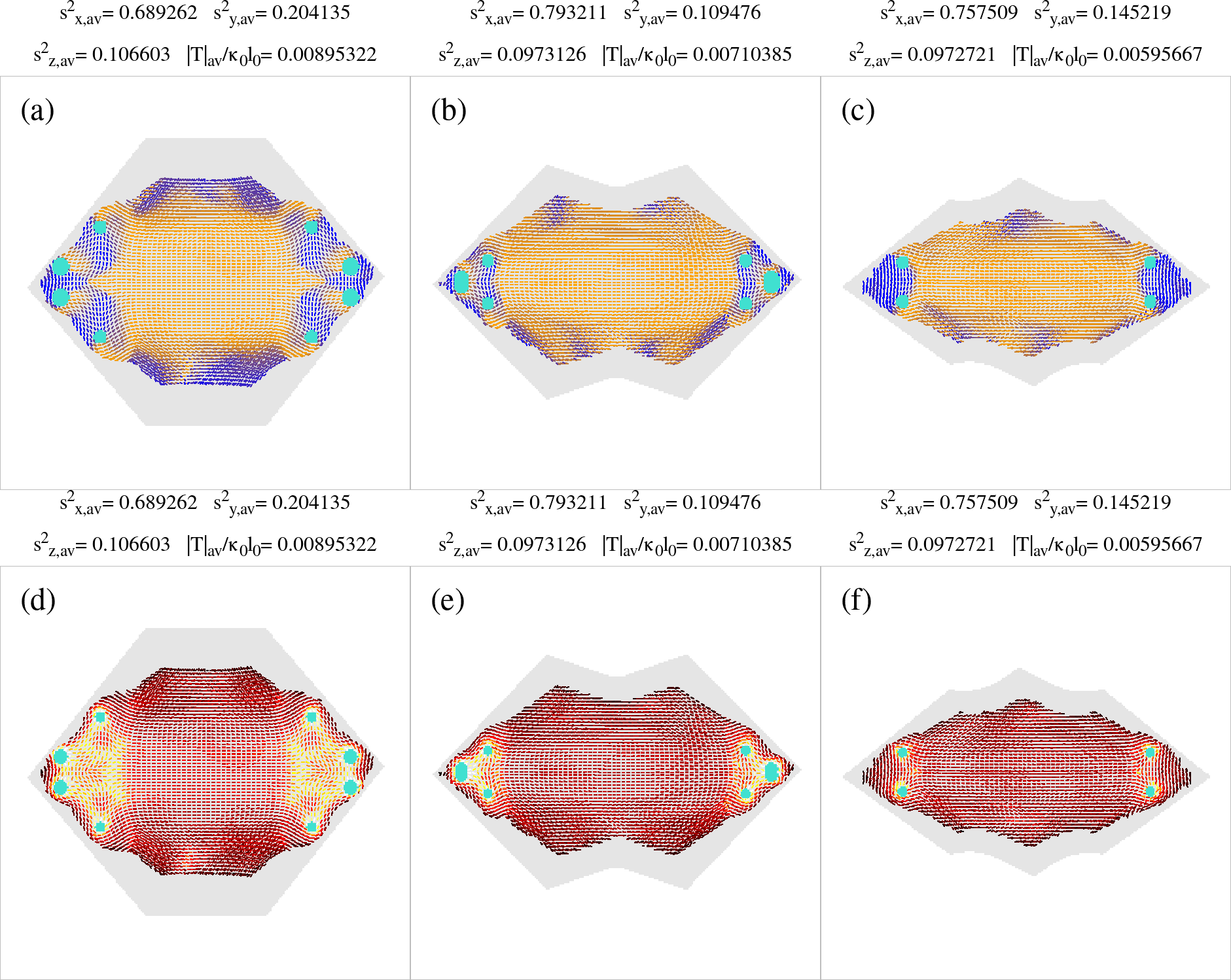}
\caption{Strategies with tethers in a single vertex instead of a short edge (diamond tethers). In Panel (a) two columns of well separated tethers lead to a region of poor alignment between the columns and as shown in Panel (d) high tension can be seen between the columns. Panel (b): This region of misalignment is minimised by moving the tethers closer. Panel (c) A single column of tethers also leads to high alignment, although the region of misalignment is larger in this case. Panels (e) and (f) show the tension corresponding to the cases in Panels (b) and (c) respectively.}
\label{fig:vertextethers}
\end{figure*}

For strategies involving moulds with a single end vertex, a single column of tethers may also be effective and reduce tension. In Fig. \ref{fig:vertextethers}, we examine tethering strategies where there is a single vertex at the end of the mould, rather than a short edge. In those cases the best moulds had two columns of tethers close to the point of the vertex (see Fig. \ref{fig:bestorient}). There are no major differences if a single tethering column is used (compare Fig. \ref{fig:vertextethers}(b) and Fig. \ref{fig:vertextethers}(c)). On the other hand, if the two columns of tethers become widely separated, misalignments occur (compare  Fig. \ref{fig:vertextethers}(a) and  Fig. \ref{fig:vertextethers}(b)). The corresponding tension per cell is shown in Fig. \ref{fig:vertextethers}(d,e,f), and is high between columns of tethers, increasing with the distance between the columns. In contrast, the tension is relatively low in the case where there is a single row of tethers.

\section{Discussion}
\label{sec:discussion}

Currently, a limited number of tethered scaffold mould designs are used for lab-grown tissues with a few common designs which all suffer from misaligned cells ($\Delta$ regions). Tendons and muscle have been grown in moulds with two tethering points (one at each end), which leads to a single aligned fibre between the tethers and a region of misalignment around the tether \cite{capel2019a,wragg2019a,garvin2003a}. The second grouping includes hydrogels  containing astrocytes which have been grown in moulds with four tethering points \cite{orourke2017}, astrocytes and corneal tissues that have been grown within an I shaped clamp where multiple tethering points are found in a line towards the end of the mould \cite{orourke2015,mukhey2018}, and moulds with attachment bars that have been used to grow fibroblast containing hydrogels \cite{eastwood1998}. In those cases, a $\Delta$ region containing unwanted misaligned tissue can be found. Highly aligned astrocytes grown in such moulds have been considered as a potential guide for neurons during nerve regeneration, and for such applications the misaligned region is a hindrance to neuron growth \cite{east2010}. 

We have shown how it is possible to use a computational model to predict which mould designs can reduce the $\Delta$ region. We have concentrated on nearly flat moulds suitable for growth of elongated cells with a preference to align e.g. fibroblasts, tendonocytes, astrocytes and cornea, although we expect the principles to work for fully three dimensional scaffolds and any kind of cells that become highly aligned within tissues. The best mould designs proposed here have potential to improve guides for neuronal growth and for use in the development of engineered tissue for assays and other applications where high alignment is required. The significance of the use of a biophysical model for mould design is twofold. Designing moulds and scaffolds that lead to specific patterns of self-organisation in artificial tissues is a slow and painstaking effort, so tools such as biophysical models offer the potential to design moulds quickly to grow specific tissue structures for e.g. pharmaceutical assays, drug screening and regenerative medicine. Also, biophysical models are applicable to the self-organisation in three-dimensional scaffolds where it is much harder to assess the organisation of lab-grown tissue. The key advantage of the biophysical model is the rate of throughput, since high performance computing can be used to make calculations for high numbers of mould designs in parallel and also the organisation and other properties of cells in those tissues, such as tension, can be immediately computed by the model without the need to make complex experimental measurements to determine those properties. Ultimately, biological experiments will be needed to confirm the predictions, but only the best candidates need to be explored, making the experimental effort targeted and as efficient as possible.

\section{Conclusions}
\label{sec:conclusions}

In summary, we have applied a microscopic biophysical model for self-organisation due to cell-matrix interactions in lab-grown polarised tissues, to the design of moulds for highly aligned cultured tissue growth. We predict such moulds would be suitable for growth of highly aligned neural tissue, muscle, tendons, cornea and fibroblasts. We demonstrated proof-of-concept for using the model for high-throughput rational design of artificial tissues. We used our biophysical model to make high-throughput calculations for a wide range of computer-generated moulds. We used alignment and tension to identify, computationally, the relative merit of moulds for growing artificial neural tissue. Using this process, we have identified candidate mould designs for growth of artificial neural and other tissue with favourable properties. To the best of our knowledge, this is the first time a biophysical model has been used for the design of tethered scaffolds or moulds.

In conclusion, two strategies were identified to increase alignment and minimise regions of misalignment: (1) indentation on the short edge of the mould, with tethers then placed within the protrusions on either side of the indentation (2) tethers inside a single vertex at the end of the mould. Precise positioning of tethers was found to be critical to optimise alignment, and small changes in tether positions can lead to enlargements in the $\Delta$-region, a decrease in alignment and an increase in tension within the tissue. The moulds are elongated with tethers located at the short edge of the tether. In the best cases the moulds widen away from the short edge. 

We briefly discuss several aspects related to the future outlook of the approach. Our goal here has been to find moulds to create tethered hydrogels (artificial or cultured tissues) with high levels of alignment and low internal tension. As such, we have shown proof-of-principle for high-throughput calculations suitable for applications requiring such high alignment. A next step will be to grow and analyse engineered tissue within these designs to confirm the predictions. For other applications it may be necessary to optimise mould designs to grow tissues with different cellular arrangements. The same design principles can be applied to other applications: quasi-random moulds with or without symmetries can be created in large numbers and then high-throughput simulation of these moulds using high-performance computing can be used to rank the solutions. Further work will include model extensions that will lead to further tools for rational design of moulds, scaffolds, and other approaches for growing artificial tissue. Designs could have multiple applications, including pharmaceutical assays, regenerative medicine and to obtain appropriate texturing within cultured meat. We will also work on extensions for the design of 3D scaffolds. Finally, we are exploring the application of deep learning to the prediction of self-organisation in tissues, and have recently developed a machine learning approach based on CONDOR for rapid prediction of the arrangements of cells and the tensions between them in tissues \cite{andrews2023}.

\section*{Acknowledgements}

We acknowledge funding from the STFC Impact Acceleration Account 2021-2022. We thank James Philips for useful discussions. We have registered moulds based on the cases shown in Fig. 3, Panels (a) through (i) and Fig. 6(i) as designs with UK registered design numbers: 6275900,  6275896, 6275894, 6275883, 6275867, 6275866, 6275865, 6275862, 6275860 and 6275851.

\bibliography{references}

\begin{thebibliography}{17}%
\makeatletter
\providecommand \@ifxundefined [1]{%
 \@ifx{#1\undefined}
}%
\providecommand \@ifnum [1]{%
 \ifnum #1\expandafter \@firstoftwo
 \else \expandafter \@secondoftwo
 \fi
}%
\providecommand \@ifx [1]{%
 \ifx #1\expandafter \@firstoftwo
 \else \expandafter \@secondoftwo
 \fi
}%
\providecommand \natexlab [1]{#1}%
\providecommand \enquote  [1]{``#1''}%
\providecommand \bibnamefont  [1]{#1}%
\providecommand \bibfnamefont [1]{#1}%
\providecommand \citenamefont [1]{#1}%
\providecommand \href@noop [0]{\@secondoftwo}%
\providecommand \href [0]{\begingroup \@sanitize@url \@href}%
\providecommand \@href[1]{\@@startlink{#1}\@@href}%
\providecommand \@@href[1]{\endgroup#1\@@endlink}%
\providecommand \@sanitize@url [0]{\catcode `\\12\catcode `\$12\catcode
  `\&12\catcode `\#12\catcode `\^12\catcode `\_12\catcode `\%12\relax}%
\providecommand \@@startlink[1]{}%
\providecommand \@@endlink[0]{}%
\providecommand \url  [0]{\begingroup\@sanitize@url \@url }%
\providecommand \@url [1]{\endgroup\@href {#1}{\urlprefix }}%
\providecommand \urlprefix  [0]{URL }%
\providecommand \Eprint [0]{\href }%
\providecommand \doibase [0]{https://doi.org/}%
\providecommand \selectlanguage [0]{\@gobble}%
\providecommand \bibinfo  [0]{\@secondoftwo}%
\providecommand \bibfield  [0]{\@secondoftwo}%
\providecommand \translation [1]{[#1]}%
\providecommand \BibitemOpen [0]{}%
\providecommand \bibitemStop [0]{}%
\providecommand \bibitemNoStop [0]{.\EOS\space}%
\providecommand \EOS [0]{\spacefactor3000\relax}%
\providecommand \BibitemShut  [1]{\csname bibitem#1\endcsname}%
\let\auto@bib@innerbib\@empty
\bibitem [{\citenamefont {Bajaj}\ \emph {et~al.}(2014)\citenamefont {Bajaj},
  \citenamefont {Schweller}, \citenamefont {Khademhosseini}, \citenamefont
  {West},\ and\ \citenamefont {Bashir}}]{bajaj2014a}%
  \BibitemOpen
  \bibfield  {author} {\bibinfo {author} {\bibfnamefont {P.}~\bibnamefont
  {Bajaj}}, \bibinfo {author} {\bibfnamefont {R.~M.}\ \bibnamefont
  {Schweller}}, \bibinfo {author} {\bibfnamefont {A.}~\bibnamefont
  {Khademhosseini}}, \bibinfo {author} {\bibfnamefont {J.~L.}\ \bibnamefont
  {West}},\ and\ \bibinfo {author} {\bibfnamefont {R.}~\bibnamefont {Bashir}},\
  }\bibfield  {title} {\bibinfo {title} {{3D} biofabrication strategies for
  tissue engineering and regenerative medicine},\ }\href@noop {} {\bibfield
  {journal} {\bibinfo  {journal} {Annu. Rev. Biomed. Eng.}\ }\textbf {\bibinfo
  {volume} {16}},\ \bibinfo {pages} {247} (\bibinfo {year} {2014})}\BibitemShut
  {NoStop}%
\bibitem [{\citenamefont {Weinhart}\ \emph {et~al.}(2019)\citenamefont
  {Weinhart}, \citenamefont {Hocke}, \citenamefont {Hippenstiel}, \citenamefont
  {Kurreck},\ and\ \citenamefont {Hedtrich}}]{weinhart2019a}%
  \BibitemOpen
  \bibfield  {author} {\bibinfo {author} {\bibfnamefont {M.}~\bibnamefont
  {Weinhart}}, \bibinfo {author} {\bibfnamefont {A.}~\bibnamefont {Hocke}},
  \bibinfo {author} {\bibfnamefont {S.}~\bibnamefont {Hippenstiel}}, \bibinfo
  {author} {\bibfnamefont {J.}~\bibnamefont {Kurreck}},\ and\ \bibinfo {author}
  {\bibfnamefont {S.}~\bibnamefont {Hedtrich}},\ }\bibfield  {title} {\bibinfo
  {title} {{3D} organ models - revolution in pharmacological research?},\
  }\href {https://www.sciencedirect.com/science/article/pii/S1043661818313574}
  {\bibfield  {journal} {\bibinfo  {journal} {Pharmacol. Res.}\ }\textbf
  {\bibinfo {volume} {139}},\ \bibinfo {pages} {446} (\bibinfo {year}
  {2019})}\BibitemShut {NoStop}%
\bibitem [{\citenamefont {Jensen}\ \emph {et~al.}(2018)\citenamefont {Jensen},
  \citenamefont {Morrill},\ and\ \citenamefont {Huang}}]{jensen2018a}%
  \BibitemOpen
  \bibfield  {author} {\bibinfo {author} {\bibfnamefont {G.}~\bibnamefont
  {Jensen}}, \bibinfo {author} {\bibfnamefont {C.}~\bibnamefont {Morrill}},\
  and\ \bibinfo {author} {\bibfnamefont {Y.}~\bibnamefont {Huang}},\ }\bibfield
   {title} {\bibinfo {title} {3{D} tissue engineering, an emerging technique
  for pharmaceutical research},\ }\href
  {https://www.sciencedirect.com/science/article/pii/S2211383517305464}
  {\bibfield  {journal} {\bibinfo  {journal} {Acta Pharm. Sin. B.}\ }\textbf
  {\bibinfo {volume} {8}},\ \bibinfo {pages} {756} (\bibinfo {year}
  {2018})}\BibitemShut {NoStop}%
\bibitem [{\citenamefont {Ben-Arye}\ and\ \citenamefont
  {Levenberg}(2019)}]{benarye2019a}%
  \BibitemOpen
  \bibfield  {author} {\bibinfo {author} {\bibfnamefont {T.}~\bibnamefont
  {Ben-Arye}}\ and\ \bibinfo {author} {\bibfnamefont {S.}~\bibnamefont
  {Levenberg}},\ }\bibfield  {title} {\bibinfo {title} {Tissue engineering for
  clean meat production},\ }\href@noop {} {\bibfield  {journal} {\bibinfo
  {journal} {Front. Sustain. Food Syst.}\ }\textbf {\bibinfo {volume} {3}},\
  \bibinfo {pages} {00046} (\bibinfo {year} {2019})}\BibitemShut {NoStop}%
\bibitem [{\citenamefont {Capel}\ \emph {et~al.}(2019)\citenamefont {Capel},
  \citenamefont {Rimington}, \citenamefont {Fleming}, \citenamefont {Player},
  \citenamefont {Baker}, \citenamefont {Turner}, \citenamefont {Jones},
  \citenamefont {Martin}, \citenamefont {Ferguson}, \citenamefont {Mudera},\
  and\ \citenamefont {Lewis}}]{capel2019a}%
  \BibitemOpen
  \bibfield  {author} {\bibinfo {author} {\bibfnamefont {A.~J.}\ \bibnamefont
  {Capel}}, \bibinfo {author} {\bibfnamefont {R.~P.}\ \bibnamefont
  {Rimington}}, \bibinfo {author} {\bibfnamefont {J.~W.}\ \bibnamefont
  {Fleming}}, \bibinfo {author} {\bibfnamefont {D.~J.}\ \bibnamefont {Player}},
  \bibinfo {author} {\bibfnamefont {L.~A.}\ \bibnamefont {Baker}}, \bibinfo
  {author} {\bibfnamefont {M.~C.}\ \bibnamefont {Turner}}, \bibinfo {author}
  {\bibfnamefont {J.~M.}\ \bibnamefont {Jones}}, \bibinfo {author}
  {\bibfnamefont {N.~R.~W.}\ \bibnamefont {Martin}}, \bibinfo {author}
  {\bibfnamefont {R.~A.}\ \bibnamefont {Ferguson}}, \bibinfo {author}
  {\bibfnamefont {V.~C.}\ \bibnamefont {Mudera}},\ and\ \bibinfo {author}
  {\bibfnamefont {M.~P.}\ \bibnamefont {Lewis}},\ }\bibfield  {title} {\bibinfo
  {title} {Scalable {3D} printed molds for human tissue engineered skeletal
  muscle},\ }\href@noop {} {\bibfield  {journal} {\bibinfo  {journal} {Front.
  Bioeng. Biotechnol.}\ }\textbf {\bibinfo {volume} {7}},\ \bibinfo {pages}
  {00020} (\bibinfo {year} {2019})}\BibitemShut {NoStop}%
\bibitem [{\citenamefont {Wragg}\ \emph {et~al.}(2019)\citenamefont {Wragg},
  \citenamefont {Player}, \citenamefont {Martin}, \citenamefont {Liu},\ and\
  \citenamefont {Lewis}}]{wragg2019a}%
  \BibitemOpen
  \bibfield  {author} {\bibinfo {author} {\bibfnamefont {N.~M.}\ \bibnamefont
  {Wragg}}, \bibinfo {author} {\bibfnamefont {D.~J.}\ \bibnamefont {Player}},
  \bibinfo {author} {\bibfnamefont {N.~R.~W.}\ \bibnamefont {Martin}}, \bibinfo
  {author} {\bibfnamefont {Y.}~\bibnamefont {Liu}},\ and\ \bibinfo {author}
  {\bibfnamefont {M.~P.}\ \bibnamefont {Lewis}},\ }\bibfield  {title} {\bibinfo
  {title} {Development of tissue-engineered skeletal muscle manufacturing
  variables},\ }\href@noop {} {\bibfield  {journal} {\bibinfo  {journal}
  {Biotechnol. Bioeng.}\ }\textbf {\bibinfo {volume} {116}},\ \bibinfo {pages}
  {2364} (\bibinfo {year} {2019})}\BibitemShut {NoStop}%
\bibitem [{\citenamefont {Garvin}\ \emph {et~al.}(2003)\citenamefont {Garvin},
  \citenamefont {Qi}, \citenamefont {Maloney},\ and\ \citenamefont
  {Banes}}]{garvin2003a}%
  \BibitemOpen
  \bibfield  {author} {\bibinfo {author} {\bibfnamefont {J.}~\bibnamefont
  {Garvin}}, \bibinfo {author} {\bibfnamefont {J.}~\bibnamefont {Qi}}, \bibinfo
  {author} {\bibfnamefont {M.}~\bibnamefont {Maloney}},\ and\ \bibinfo {author}
  {\bibfnamefont {A.~J.}\ \bibnamefont {Banes}},\ }\bibfield  {title} {\bibinfo
  {title} {Novel system for engineering bioartificial tendons and application
  of mechanical load},\ }\href@noop {} {\bibfield  {journal} {\bibinfo
  {journal} {Tissue Eng. Part A}\ }\textbf {\bibinfo {volume} {9}},\ \bibinfo
  {pages} {967} (\bibinfo {year} {2003})}\BibitemShut {NoStop}%
\bibitem [{\citenamefont {Georgiou}\ \emph {et~al.}(2013)\citenamefont
  {Georgiou}, \citenamefont {Bunting}, \citenamefont {Davies}, \citenamefont
  {Loughlin}, \citenamefont {Golding},\ and\ \citenamefont
  {Phillips}}]{georgiou2013}%
  \BibitemOpen
  \bibfield  {author} {\bibinfo {author} {\bibfnamefont {M.}~\bibnamefont
  {Georgiou}}, \bibinfo {author} {\bibfnamefont {S.~C.}\ \bibnamefont
  {Bunting}}, \bibinfo {author} {\bibfnamefont {H.~A.}\ \bibnamefont {Davies}},
  \bibinfo {author} {\bibfnamefont {A.~J.}\ \bibnamefont {Loughlin}}, \bibinfo
  {author} {\bibfnamefont {J.~P.}\ \bibnamefont {Golding}},\ and\ \bibinfo
  {author} {\bibfnamefont {J.~B.}\ \bibnamefont {Phillips}},\ }\bibfield
  {title} {\bibinfo {title} {Engineered neural tissue for peripheral nerve
  repair},\ }\href@noop {} {\bibfield  {journal} {\bibinfo  {journal}
  {Biomater.}\ }\textbf {\bibinfo {volume} {34}},\ \bibinfo {pages} {7335}
  (\bibinfo {year} {2013})}\BibitemShut {NoStop}%
\bibitem [{\citenamefont {Mukhey}\ \emph {et~al.}(2018)\citenamefont {Mukhey},
  \citenamefont {Phillips}, \citenamefont {Daniels},\ and\ \citenamefont
  {Kureshi}}]{mukhey2018}%
  \BibitemOpen
  \bibfield  {author} {\bibinfo {author} {\bibfnamefont {D.}~\bibnamefont
  {Mukhey}}, \bibinfo {author} {\bibfnamefont {J.}~\bibnamefont {Phillips}},
  \bibinfo {author} {\bibfnamefont {J.}~\bibnamefont {Daniels}},\ and\ \bibinfo
  {author} {\bibfnamefont {A.}~\bibnamefont {Kureshi}},\ }\bibfield  {title}
  {\bibinfo {title} {Controlling human corneal stromal stem cell contraction to
  mediate rapid cell and matrix organization of real architecture for
  3-dimensional tissue equivalents},\ }\href@noop {} {\bibfield  {journal}
  {\bibinfo  {journal} {Acta Biomater.}\ }\textbf {\bibinfo {volume} {67}},\
  \bibinfo {pages} {229} (\bibinfo {year} {2018})}\BibitemShut {NoStop}%
\bibitem [{\citenamefont {Eastwood}\ \emph {et~al.}(1998)\citenamefont
  {Eastwood}, \citenamefont {Mudera}, \citenamefont {McGrouther},\ and\
  \citenamefont {Brown}}]{eastwood1998}%
  \BibitemOpen
  \bibfield  {author} {\bibinfo {author} {\bibfnamefont {M.}~\bibnamefont
  {Eastwood}}, \bibinfo {author} {\bibfnamefont {V.}~\bibnamefont {Mudera}},
  \bibinfo {author} {\bibfnamefont {D.}~\bibnamefont {McGrouther}},\ and\
  \bibinfo {author} {\bibfnamefont {R.}~\bibnamefont {Brown}},\ }\bibfield
  {title} {\bibinfo {title} {Effect of precise mechanical loading on fibroblast
  populated collagen lattices: Morphological changes},\ }\href@noop {}
  {\bibfield  {journal} {\bibinfo  {journal} {Cell Motil. Cytoskeleton}\
  }\textbf {\bibinfo {volume} {40}},\ \bibinfo {pages} {13–21} (\bibinfo
  {year} {1998})}\BibitemShut {NoStop}%
\bibitem [{\citenamefont {Hague}\ \emph {et~al.}(2020)\citenamefont {Hague},
  \citenamefont {Mieczkowski}, \citenamefont {O’Rourke}, \citenamefont
  {Loughlin},\ and\ \citenamefont {Phillips}}]{hague2019a}%
  \BibitemOpen
  \bibfield  {author} {\bibinfo {author} {\bibfnamefont {J.~P.}\ \bibnamefont
  {Hague}}, \bibinfo {author} {\bibfnamefont {P.~W.}\ \bibnamefont
  {Mieczkowski}}, \bibinfo {author} {\bibfnamefont {C.}~\bibnamefont
  {O’Rourke}}, \bibinfo {author} {\bibfnamefont {A.~J.}\ \bibnamefont
  {Loughlin}},\ and\ \bibinfo {author} {\bibfnamefont {J.~B.}\ \bibnamefont
  {Phillips}},\ }\bibfield  {title} {\bibinfo {title} {Microscopic biophysical
  model of self-organization in tissue due to feedback between cell- and
  macroscopic-scale forces},\ }\href@noop {} {\bibfield  {journal} {\bibinfo
  {journal} {Phys. Rev. Res.}\ }\textbf {\bibinfo {volume} {2}},\ \bibinfo
  {pages} {043217} (\bibinfo {year} {2020})}\BibitemShut {NoStop}%
\bibitem [{\citenamefont {Andrews}\ \emph {et~al.}(2023)\citenamefont
  {Andrews}, \citenamefont {H.Dickinson},\ and\ \citenamefont
  {Hague}}]{andrews2023}%
  \BibitemOpen
  \bibfield  {author} {\bibinfo {author} {\bibfnamefont {A.}~\bibnamefont
  {Andrews}}, \bibinfo {author} {\bibnamefont {H.Dickinson}},\ and\ \bibinfo
  {author} {\bibfnamefont {J.}~\bibnamefont {Hague}},\ }\href@noop {}
  {\bibfield  {journal} {\bibinfo  {journal} {arXiv:}\ ,\ \bibinfo {pages}
  {2303.18017}} (\bibinfo {year} {2023})}\BibitemShut {NoStop}%
\bibitem [{\citenamefont {Kular}\ \emph {et~al.}(2014)\citenamefont {Kular},
  \citenamefont {Basu},\ and\ \citenamefont {Sharma}}]{kular2014a}%
  \BibitemOpen
  \bibfield  {author} {\bibinfo {author} {\bibfnamefont {J.~K.}\ \bibnamefont
  {Kular}}, \bibinfo {author} {\bibfnamefont {S.}~\bibnamefont {Basu}},\ and\
  \bibinfo {author} {\bibfnamefont {R.~I.}\ \bibnamefont {Sharma}},\ }\bibfield
   {title} {\bibinfo {title} {The extracellular matrix: Structure, composition,
  age-related differences, tools for analysis and applications for tissue
  engineering},\ }\href@noop {} {\bibfield  {journal} {\bibinfo  {journal} {J.
  Tissue Eng.}\ }\textbf {\bibinfo {volume} {5}},\ \bibinfo {pages}
  {2041731414557112} (\bibinfo {year} {2014})}\BibitemShut {NoStop}%
\bibitem [{\citenamefont {Gillies}\ \emph {et~al.}()\citenamefont {Gillies},
  \citenamefont {van~der Wel}, \citenamefont {van~den Bossche}, \citenamefont
  {Taves}, \citenamefont {Arnott},\ and\ \citenamefont {Ward}}]{shapely2023}%
  \BibitemOpen
  \bibfield  {author} {\bibinfo {author} {\bibfnamefont {S.}~\bibnamefont
  {Gillies}}, \bibinfo {author} {\bibfnamefont {C.}~\bibnamefont {van~der
  Wel}}, \bibinfo {author} {\bibfnamefont {J.}~\bibnamefont {van~den Bossche}},
  \bibinfo {author} {\bibfnamefont {M.}~\bibnamefont {Taves}}, \bibinfo
  {author} {\bibfnamefont {J.}~\bibnamefont {Arnott}},\ and\ \bibinfo {author}
  {\bibfnamefont {B.}~\bibnamefont {Ward}},\ }\href@noop {} {\bibinfo {title}
  {Shapely}},\ \bibinfo {note} {version 2.0.1 (Jan 30th 2023),
  doi:10.5281/zenodo.5597138, https://pypi.org/project/Shapely and
  https://github.com/shapely/shapely}\BibitemShut {NoStop}%
\bibitem [{\citenamefont {O’Rourke}\ \emph {et~al.}(2017)\citenamefont
  {O’Rourke}, \citenamefont {Lee-Reeves}, \citenamefont {Drake},
  \citenamefont {Cameron}, \citenamefont {Loughlin},\ and\ \citenamefont
  {Phillips}}]{orourke2017}%
  \BibitemOpen
  \bibfield  {author} {\bibinfo {author} {\bibfnamefont {C.}~\bibnamefont
  {O’Rourke}}, \bibinfo {author} {\bibfnamefont {C.}~\bibnamefont
  {Lee-Reeves}}, \bibinfo {author} {\bibfnamefont {R.~A.~L.}\ \bibnamefont
  {Drake}}, \bibinfo {author} {\bibfnamefont {G.~W.~W.}\ \bibnamefont
  {Cameron}}, \bibinfo {author} {\bibfnamefont {A.~J.}\ \bibnamefont
  {Loughlin}},\ and\ \bibinfo {author} {\bibfnamefont {J.~B.}\ \bibnamefont
  {Phillips}},\ }\bibfield  {title} {\bibinfo {title} {Adapting
  tissue-engineered in vitro cns models for high-throughput study of
  neurodegeneration},\ }\href@noop {} {\bibfield  {journal} {\bibinfo
  {journal} {J. Tissue Eng.}\ }\textbf {\bibinfo {volume} {8}},\ \bibinfo
  {pages} {1} (\bibinfo {year} {2017})}\BibitemShut {NoStop}%
\bibitem [{\citenamefont {O'Rourke}\ \emph {et~al.}(2015)\citenamefont
  {O'Rourke}, \citenamefont {Drake}, \citenamefont {Cameron}, \citenamefont
  {Loughlin},\ and\ \citenamefont {Phillips}}]{orourke2015}%
  \BibitemOpen
  \bibfield  {author} {\bibinfo {author} {\bibfnamefont {C.}~\bibnamefont
  {O'Rourke}}, \bibinfo {author} {\bibfnamefont {R.~A.}\ \bibnamefont {Drake}},
  \bibinfo {author} {\bibfnamefont {G.~W.}\ \bibnamefont {Cameron}}, \bibinfo
  {author} {\bibfnamefont {J.~A.}\ \bibnamefont {Loughlin}},\ and\ \bibinfo
  {author} {\bibfnamefont {J.~B.}\ \bibnamefont {Phillips}},\ }\bibfield
  {title} {\bibinfo {title} {Optimising contraction and alignment of cellular
  collagen hydrogels to achieve reliable and consistent engineered anisotropic
  tissue.},\ }\href@noop {} {\bibfield  {journal} {\bibinfo  {journal} {J.
  Biomater. Appl.}\ }\textbf {\bibinfo {volume} {30}},\ \bibinfo {pages} {599}
  (\bibinfo {year} {2015})}\BibitemShut {NoStop}%
\bibitem [{\citenamefont {East}\ \emph {et~al.}(2010)\citenamefont {East},
  \citenamefont {De~Oliviera}, \citenamefont {Golding},\ and\ \citenamefont
  {Phillips}}]{east2010}%
  \BibitemOpen
  \bibfield  {author} {\bibinfo {author} {\bibfnamefont {E.}~\bibnamefont
  {East}}, \bibinfo {author} {\bibfnamefont {D.~B.}\ \bibnamefont
  {De~Oliviera}}, \bibinfo {author} {\bibfnamefont {J.~P.}\ \bibnamefont
  {Golding}},\ and\ \bibinfo {author} {\bibfnamefont {J.~B.}\ \bibnamefont
  {Phillips}},\ }\bibfield  {title} {\bibinfo {title} {Alignment of astrocytes
  increases neuronal growth in three-dimensional collagen gels and is
  maintained following plastic compression to form a spinal cord repair
  conduit},\ }\href@noop {} {\bibfield  {journal} {\bibinfo  {journal} {Tissue
  Eng. Part A}\ }\textbf {\bibinfo {volume} {16}},\ \bibinfo {pages} {3173}
  (\bibinfo {year} {2010})}\BibitemShut {NoStop}%
\end{thebibliography}%

\end{document}